\documentstyle[12pt,times,epsf]{article}

\begin{document}

\baselineskip 6mm
\renewcommand{\thefootnote}{\fnsymbol{footnote}}


\newcommand{\nc}{\newcommand}
\newcommand{\rnc}{\renewcommand}


\rnc{\baselinestretch}{1.24}	
\setlength{\jot}{6pt} 		
\rnc{\arraystretch}{1.24}   	

\makeatletter
\rnc{\theequation}{\thesection.\arabic{equation}}
\@addtoreset{equation}{section}
\makeatother                      



\nc{\be}{\begin{equation}}
\nc{\ee}{\end{equation}}
\nc{\bea}{\begin{eqnarray}}
\nc{\eea}{\end{eqnarray}}
\nc{\xx}{\nonumber\\}

\nc{\eq}[1]{(\ref{#1})}
\nc{\newcaption}[1]{\centerline{\parbox{6in}{\caption{#1}}}}

\nc{\fig}[3]{
\begin{figure}
\centerline{\epsfxsize=#1\epsfbox{#2.eps}}
\newcaption{#3. \label{#2}}
\end{figure}
}


\nc{\np}[3]{Nucl. Phys. {\bf B#1} (#2) #3}
\nc{\pl}[3]{Phys. Lett. {\bf #1B} (#2) #3}
\nc{\prl}[3]{Phys. Rev. Lett.{\bf #1} (#2) #3}
\nc{\prd}[3]{Phys. Rev. {\bf D#1} (#2) #3}
\nc{\ap}[3]{Ann. Phys. {\bf #1} (#2) #3}
\nc{\prep}[3]{Phys. Rep. {\bf #1} (#2) #3}
\nc{\rmp}[3]{Rev. Mod. Phys. {\bf #1} (#2) #3}
\nc{\cmp}[3]{Comm. Math. Phys. {\bf #1} (#2) #3}
\nc{\mpl}[3]{Mod. Phys. Lett. {\bf #1} (#2) #3}
\nc{\cqg}[3]{Class. Quant. Grav. {\bf #1} (#2) #3}
\nc{\jhep}[3]{J. High Energy Phys. {\bf #1} (#2) #3}


\def\vare{\varepsilon}
\def\bz{\bar{z}}
\def\bw{\bar{w}}


\def\CA{{\cal A}}
\def\CC{{\cal C}}
\def\CD{{\cal D}}
\def\CE{{\cal E}}
\def\CF{{\cal F}}
\def\CG{{\cal G}}
\def\CT{{\cal T}}
\def\CM{{\cal M}}
\def\CN{{\cal N}}
\def\CP{{\cal P}}
\def\CL{{\cal L}}
\def\CV{{\cal V}}
\def\CS{{\cal S}}
\def\CW{{\cal W}}
\def\CY{{\cal Y}}
\def\CS{{\cal S}}
\def\CO{{\cal O}}
\def\CP{{\cal P}}
\def\CN{{\cal N}}


\def\IR{{\hbox{{\rm I}\kern-.2em\hbox{\rm R}}}}
\def\IB{{\hbox{{\rm I}\kern-.2em\hbox{\rm B}}}}
\def\IN{{\hbox{{\rm I}\kern-.2em\hbox{\rm N}}}}
\def\IC{\,\,{\hbox{{\rm I}\kern-.59em\hbox{\bf C}}}}
\def\IZ{{\hbox{{\rm Z}\kern-.4em\hbox{\rm Z}}}}
\def\IP{{\hbox{{\rm I}\kern-.2em\hbox{\rm P}}}}
\def\IH{{\hbox{{\rm I}\kern-.4em\hbox{\rm H}}}}
\def\ID{{\hbox{{\rm I}\kern-.2em\hbox{\rm D}}}}


\def\a{\alpha}
\def\b{\beta}
\def\ga{\gamma}
\def\d{\delta}
\def\ep{\epsilon}
\def\ph{\phi}
\def\k{\kappa}
\def\l{\lambda}
\def\m{\mu}
\def\n{\nu}
\def\th{\theta}
\def\rh{\rho}
\def\s{\sigma}
\def\t{\tau}
\def\w{\omega}
\def\G{\Gamma}


\def\half{\frac{1}{2}}
\def\imp{\Longrightarrow}
\def\dint#1#2{\int\limits_{#1}^{#2}}
\def\goto{\rightarrow}
\def\para{\parallel}
\def\brac#1{\langle #1 \rangle}
\def\del{\nabla}
\def\grad{\nabla}
\def\curl{\nabla\times}
\def\div{\nabla\cdot}
\def\p{\partial}
\def\e{\epsilon_0}


\def\Tr{{\rm Tr}}
\def\det{{\rm det}}


\def\Kahler{K\"{a}hler}


\def\e{\varepsilon}
\def\bA{\bar{A}}
\def\c{\zeta}

\begin{titlepage}
\hfill\parbox{4cm}
{hep-th/9909059 \\ KIAS-P99081 \\ UTTG-04-99}

\vspace{15mm}
\centerline{\Large \bf Closed Strings Interacting with 
 Noncommutative D-branes}
\vspace{10mm}
\begin{center} 
Seungjoon Hyun, 
Youngjai Kiem, 
Sangmin Lee\footnote{hyun, ykiem, sangmin@kias.re.kr} 
\\[2mm] 
{\sl School of Physics, Korea Institute for Advanced Study, 
Seoul 130-012, Korea}
\\[5mm]
and
\\[5mm]
Chang-Yeong Lee\footnote{leecy@zippy.ph.utexas.edu} 
\\[2mm]
{\sl Theory Group, Department of Physics, University of Texas,  
Austin, TX 78712, USA, \\
Department of Physics, Sejong Univeristy, 
Seoul 143-747, Korea}
\end{center}
\thispagestyle{empty}
\vskip 15mm

\centerline{\bf ABSTRACT}
\vskip 5mm
\noindent
Closed string dynamics in the presence of 
noncommutative D$p$-branes is investigated.
In particular, we compute bulk closed string 
two-point scattering amplitudes; the bulk
space-time geometries encoded in the amplitudes
are shown to be consistent with the recently 
proposed background space-time geometries dual 
to noncommutative Yang-Mills theories.
Three-point closed string absorption/emission
amplitudes are obtained to show some features
of noncommutative D$p$-branes, such as modified 
pole structures and exponential phase factors 
linearly proportional to the external closed 
string momentum. 
\vspace{2cm}
\end{titlepage}

\baselineskip 7mm
\renewcommand{\thefootnote}{\arabic{footnote}}
\setcounter{footnote}{0}

\section{Introduction}

In the framework of open strings, it is now possible
to systematically study the physics on noncommutative
space-time \cite{doughull}-\cite{witten}.  
In particular, a natural vantage
point for the investigation of $(p+1)$-dimensional 
gauge theories on a noncommutative space is to
consider the world-volume theory of D$p$-branes
when the constant background NS-NS two-form 
gauge fields parallel to the branes are turned on.  
One can then study open string dynamics stuck
to the branes and find that, in an appropriate
decoupling limit, the world-volume theory corresponds
to noncommutative Yang-Mills 
theory \cite{doughull}-\cite{witten}.  

The main theme of this paper is to understand
closed string dynamics in the presence of 
noncommutative D$p$-branes.  On top of an obvious
observation that any theory of open strings 
should include closed strings, there are other
reasons to study closed string dynamics in this 
context.  By now, there are considerable body
of evidence toward the validity of the duality
between the conformal field theory on a 
$(p+1)$-dimensional space and the 
string theory/supergravity on a 
$(p+2)$-dimensional Anti-de Sitter  
space \cite{adscft}.
One natural question following the understanding
of noncommutative Yang-Mills theory is to
find an appropriate dual background space-time
geometry.  Recently, such dual background 
space-time geometries were proposed by Hashimoto
and Itzhaki \cite{aki}, and Maldacena and 
Russo \cite{maldaruss}, following the chain 
of T-duality arguments \cite{doughull}.  
These background space-time geometries can be 
directly probed by closed strings moving on it.  
Thus, the study of the closed string two-point 
scattering amplitudes shown in Fig.~1 may provide
us with a direct string theoretic justification
for the proposed background geometries.  The 
second issue is to understand noncommutative
D-brane black holes.  It has been noted
that the Hawking radiation from
D-brane black holes and its time-reversal 
process, the matter absorption into such black 
holes, can be understood via the
microscopic D-brane description by 
considering three-point amplitudes of the type shown 
in Fig.~2 \cite{klebanov}.  
One then wonders what will be the
effect of the world-volume noncommutativity
on the emission/absorption processes from D-branes.
Finally, it is suggested that the noncommutative
gauge theories play a crucial role in the
development of string field 
theories \cite{witten, witten2}.  
Understanding subleading effects in $1/N$ is 
important in the construction of string field 
theories, and the inclusion of the closed strings 
is an essential step for that purpose; one needs 
to evaluate correlation functions on general 
Riemann surfaces with boundaries along with 
marked points and handles.  It
will be nice to have a simple calculational
prescription to take into account of the 
effect of the background NS-NS two-form 
gauge fields in such a context\footnote{In the case
of purely open string diagrams, a simple 
prescription in this regard is available
in Refs.~\cite{schom,witten}.}.  In this paper, we 
address the 
first two issues and, in that process, make a 
modest progress toward the third issue.  

In Section 2, after setting up our notations
based on Refs.~\cite{fradtsey,clny},
we explain how the noncommutativity effects
are implemented by simple global rotations of the 
two string coordinates parameterizing a two-cycle
along which the NS-NS two-form gauge 
field $B_{\mu \nu}$
has a non-vanishing component whose strength
determines the global rotation angle.  
Starting from the usual Neumann boundary conditions
where we identify the holomorphic sector and the
anti-holomorphic sector at the world-sheet boundary,
the operator products for the general value 
of $B_{\mu \nu}$ can thus be simply determined.
The senses of the rotation for the holomorphic 
sector and the anti-holomorphic sector are 
opposite to each other; therefore, the 
$\pi /2 $ rotation (large $B_{\mu \nu}$ limit)
flips the Neumann boundary 
condition into the Dirichlet boundary condition,
turning D$p$-branes into D$(p-2)$-branes.
For generic values of $B_{\mu \nu}$, both
D$p$-branes and D$(p-2)$-branes are 
present \cite{brwithin}.
The correlation 
functions in the presence of noncommutative D-branes
can be computed by inserting these rotation
matrices at appropriate steps of the calculations.

In section 3, we directly compute the tree-level
closed string two-point scattering 
amplitudes, shown in Fig.~1 in the presence
of noncommutative D$p$-branes.  In the $s$-channel
factorization limit, intermediate open string 
states are identified and are shown to be consistent 
with the picture of Seiberg and Witten \cite{witten}.  
On-shell,
this amplitude does not contain (due to the momentum
conservation parallel to the branes) the exponential 
phase factor used to 
define the $*$-product \cite{witten,connes}.  

In section 4, the same scattering
problems are also investigated on the background
geometries of Refs.~\cite{aki,maldaruss} via 
supergravity analysis at long distance (Fig.~2).  
The results turn out to be identical to those
obtained in Section 3 when we 
approach the leading $t$-channel pole for the 
string amplitudes,
justifying the background supergravity space-time
geometries of Refs.~\cite{aki,maldaruss}.   
   
In section 5, we compute the three-point amplitudes
shown in Fig.~3 to study the absorption/emission
from noncommutative D$p$-branes.  The key difference
between the commutative and the noncommutative
cases is the existence of an exponential phase 
factor for each factorization channel,
which is used to define the $*$-product in the 
noncommutative case \cite{witten}.  
By the momentum conservation parallel to 
the branes, the exponential phase depends 
linearly on the external closed string momentum
parallel to the branes; it vanishes
for the external closed string momentum with 
vanishing parallel 
components (see also \cite{bigasuss}).  Furthermore,
when compared to the commutative case, the
pole and zero structures of the amplitudes
are distinctively changed.  

In Section 6, we discuss the implications and
possible generalizations of our calculations
presented in this paper. 

\section{Preliminaries}

Our computation of correlation functions 
will be based on the modern covariant formulation
defined via a conformal field theory on
the open string world-sheet.  Our primary
interest is to understand how the existence
of the background NS-NS two-form gauge 
field affects the correlation functions
involving closed strings. In the context
of pure open strings, this issue was 
investigated in \cite{schom,witten}. 
A useful starting point for this purpose is to consider
the open string action
\be
S = -\frac{1}{4 \pi \a'} \int d^2 \s ( 
g_{\m\n} \partial^a X^\m \partial_a X^\n 
+ i B_{\mu \nu} \ep^{ab} \partial_a X^\m \partial_b X^\n) + 
\mbox{fermionic part},
\label{2dws}
\ee
where $X^\m$, $g_{\m\n}$ and $B_{\m\n}$ 
represent the string coordinates, the constant background metric and 
the constant background NS-NS two-form field, respectively. 
The background $B$ field has non-zero components only along the
directions parallel to the D$p$-branes. 
The components perpendicular to the branes can be gauged away.

Since the $B$ field vanishes along
the directions perpendicular to the 
branes, we can simply impose the usual
Dirichlet boundary conditions for the
perpendicular directions.  The parallel
directions, however, even if the bulk
equations of motion are not affected
by the constant $B$ field,
the Neumann boundary condition changes to 
\be
g_{\m\n}\partial_n X^\n + i B_{\m\n} \partial_t X^\n
 = 0
\label{nbc}
\ee
on the open string end points.  Here
$\partial_n$ and $\partial_t$ denote
the normal derivative and the tangential
derivative to the boundary of the string world sheet, respectively.    

In this paper, we will be interested in disk diagrams only. 
For the computational
convenience, we map the disk to the upper
half plane, putting boundary at the real-axis on the 
complex plane. Also for simplicity, we set $g_{\m\n} = \eta_{\m\n}$
\footnote{When writing down 
products of tensor objects, especially from
Section 3, the dot product with
respect to $g^{\mu \nu}$ is implied, unless otherwise
noted.}.
It is straightforward to relax this restriction.

A useful trick to handle
the mode expansion under the boundary condition
Eq. \eq{nbc} is the following. 
Let $r=2k$ be the rank of the background $B$ field. 
Using the $SO(1,p)$ symmetry of the brane and 
redefining some coordinates, we can set 
$B_{\m\n} = 0$ for $0 \le \m \le p-r$ or $0 \le \n \le p-r$ and 
bring the remaining $r\times r$ matrix to the block diagonal form.
If we denote the $r$ coordinates by $y_i$ ($1\le i  \le r$), 
the restriction of $B$ field to $(y_{2i-1}, y_{2i})$ subspace takes the form
\be
B_i = \pmatrix{
0 &   B_i \cr
-B_i & 0}.
\ee
Introduce a matrix $R$ whose $i$-th $2\times 2$ block is 
\be
R_i = \pmatrix{
 \cos\th_i & \sin\th_i \cr
-\sin\th_i & \cos\th_i 
} ~, 
\ee      
where $\th_i \equiv \tan^{-1} B_i$, and otherwise 
equal to the identity matrix
\footnote{
The authors of Ref.~\cite{gkp} considered string scattering amplitude 
in the presence of {\em electric} flux on $D$-branes. 
They introduced a boost matrix as a function of the electric field, 
whose magnetic counterpart is our rotation matrix. 
They also observed phase factors analogous to ours \eq{t3}.
In addition, after the initial submission of this paper,
M.R. Garousi informed us that the large part of the 
calculations presented
in Sections 2 and 3 were already reported in 
Ref.~\cite{garousi}, albeit in a different language.}. 
In terms of the matrix $R$, 
the boundary condition \eq{nbc} can be rewritten as
\be
R^T \partial_z X - R \partial_{\bar{z}} X = 0,
\label{nbc1}
\ee
where we consider $X^\m$ as a column vector. 

Let $X(z)$ be the operator defined in terms of the mode expansion
\footnote{From here on, we set $\a' = 2$.}
\be
\partial X^\m(z) = -i \sum_{n=-\infty}^\infty \a_n^\m z^{-n-1} ~.
\ee
When $B_{\m\n} = 0$, 
we define the anti-holomorphic counterpart $X(\bz)$ using the {\em same}
oscillator modes $\a^\m_n$ in order to ensure the Neumann boundary condition 
on $X(z,\bz) = X(z) + X(\bz)$. 
When $B_{\m\n} \ne 0$, we note that the field  
\be
X(z,\bz) = R X(z) + R^T X (\bar{z})
\label{trans}
\ee
satisfies \eq{nbc1}.
Using the standard operator product 
\be
\langle X^\m (z) X^\n (w) \rangle = - \eta^{\m\n} \ln (z-w),
\ee
one can easily determine the operator products
for the ``rotated'' $X, \bar{X}$ fields:
\bea
\langle (RX)^\m (z) (RX)^\n (w) \rangle &=& - \eta^{\m\n} \ln (z-w),\\
\langle (R^T\bar{X})^\m (\bz) (R^T\bar{X})^\n (\bw) \rangle
&=& - \eta^{\m\n} \ln (\bz-\bw),\\
\langle (RX)^\m (z) (R^T\bar{X})^\n \rangle &=& 
-(2G^{\m\n} - \eta^{\m\n} + 2\Theta^{\m\n} )\ln (z -\bw),
\eea
where we defined, following \cite{witten}, $G^{\m\n}$ and $\Theta^{\m\n}$ as 
the symmetric and anti-symmetric part of $(\eta_{\m\n} +B_{\m\n})^{-1}$, 
respectively. Clearly, one can combine the above formulas to 
obtain the expression for the operator product of two open string vertex 
operators, originally derived in \cite{clny}

The holomorphic part and the anti-holomorphic part of the
string coordinates are
related to each other via
\begin{equation}
 \left( RX (z) \right) = R^2  \left( R^T \bar{X} (\bar{z} )
 \right)
\end{equation}
at the boundary $z = \bar{z}$.  Therefore, as $B_i \rightarrow
\infty$, i.e., $\theta_i \rightarrow \pi / 2$, the Neumann
boundary condition at $\theta_i = 0$ 
flips to the Dirichlet boundary condition,
$R_i^2 = - I$.
In the intermediate case, the boundary conditions are mixed,
$\tan \theta_i$ being the ``measure of the relative
proportion'' of the Dirichlet boundary parts 
(``D$(p-2)$-branes'')
to the Neumann boundary parts (``D$p$-branes'').


\section{Scatterings from noncommutative D-branes}

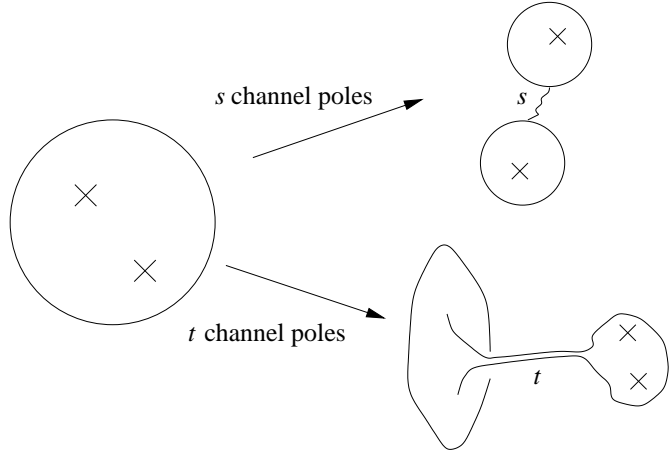
\fig{250pt}{2pt}{Two-point closed string diagram.
On the right hand side, we show a $t$-channel and
an $s$-channel factorization limit}

We now perform the calculation of the bulk two-point 
closed string amplitudes shown in Fig.~1.  After a brief 
review of the same calculation without the background
$B$ field \cite{klebanov}, we turn on the 
constant background $B$ field. 

\subsection{Review: Scatterings from commutative D-branes}

The closed string scattering amplitude from a commutative
D$p$-brane is given by 
\be
\label{twoamp}
A = \int d^2 z_1 d^2 z_2 \langle V_1(z_1,\bz_1) V_2(z_2,\bz_2) \rangle,
\ee
corresponding to the string diagram Fig.~1.  The
appropriate vertex operators are
\bea
V_1(z_1,\bz_1) &=& (\e_1 D)_{\m\n} 
   V_{-1}^\m(p_1,z_1) V_{-1}^\n(Dp_1, \bz_1) ~,
\\
V_2(z_2,\bz_2) &=& (\e_2 D)_{\m\n} 
    V_0^\m(p_2,z_2) V_0^\n (Dp_2, \bz_2) ~,
\\
\label{ghost1}
V_{-1}^\m(p_1,z_1) &=& e^{-\phi(z_1)} 
   \psi^\m(z_1) e^{ip_1 X(z_1)} ~, \\
\label{ghost0}
V_0^\m(p_2,z_2) &=& 
\{ \partial X^\m(z_2) 
 + ip_2\cdot\psi(z_2)\psi^\m(z_2) \} e^{ip_2 X(z_2)} ~, 
\eea
where the matrix $D$ is defined as
\begin{equation}
D^\m_\n = {\rm diag}
(\underbrace{+,\cdots,+}_{p+1},\underbrace{-,\cdots,-}_{9-p}) ~.
\end{equation}
The matrix $D$ is included to account for 
the Dirichlet boundary condition 
on the world-sheet field associated with the 
directions perpendicular to the brane.
Fixing the $SL(2,\IR)$ invariance of the 
amplitude \eq{twoamp}, 
and performing the remaining integral, one finds that 
\bea
A = {\G(s)\G(t)\over \G(1+s+t)} (s a_1 - t a_2),
\eea
where the two kinematic invariants are defined by
\be
s = 2p_{1\parallel}^2 = 
 2p_{2\parallel}^2,\  \ t = p_1\cdot p_2 ~,
\ee
and 
\be
\begin{array}{rcl}
a_1 &=& \Tr(\e_1 D) p_1\e_2p_1 - p_1\e_2 D \e_1 p_2 
- p_1 \e_2 \e_1^T D p_1 - p_1 \e_2^T 
\e_1 D p_1 + \{ 1 \leftrightarrow 2 \}\cr
&&-p_1\e_2\e_1^T p_2 - p_1 \e_2^T \e_1 p_2 
  - s \Tr(\e_1\e_2^T) ~, 
\cr
a_2 &=& \Tr(\e_1D)(p_1\e_2Dp_2 + p_2D\e_2 p_1 
   + p_2 D \e_2 D p_2) 
+p_1D\e_1 D \e_2 D p_2 
  + \{ 1 \leftrightarrow 2 \} \cr
&& + p_1D\e_1\e_2^TDp_2 + p_1D\e_1^T\e_2Dp_2 
+ s\Tr(\e_1D\e_2D) -s\Tr(\e_1\e_2^T) \cr
&& - (s+t) \Tr(\e_1D)\Tr(\e_2D) ~. 
\label{polar1}
\end{array}
\ee
This string amplitude\footnote{We note that in Ref.~\cite{klebanov},
there is a typographical error that appears in the expression
for $a_1$.  For the scattering processes involving $B$,
the correct formula shown here is important for the comparison
with supergravity.} is consistent with the well-known
D$p$-brane supergravity background geometry as shown
in \cite{klebanov}.  We note that the external closed string
momenta are conserved only along the directions parallel
to the branes, $p_{1 \parallel} + p_{2 \parallel} = 0$.

\subsection{Turning on the $B$ field}

As we observed in section 2, the effect of the 
constant $B$ field background 
can be incorporated simply by rotating the world-sheet fields by 
the matrix $R$ defined there. Equivalently, we can rotate the 
polarization tensors and momenta in the definitions of the vertex operators
\bea
V_1(z_1,\bz_1) &=& 
(R^T \e_1 D R^T)_{\m\n} V_{-1}^\m(R^T p_1,z_1) 
 V_{-1}^\n(RDp_1, \bz_1) ~,
\label{svert1}
\\
V_2(z_2,\bz_2) &=& 
(R^T \e_2 D R^T)_{\m\n} V_0^\m(R^T p_2,z_2) 
 V_0^\n (RDp_2, \bz_2) ~,
\label{svert2}
\eea
while leaving the definitions \eq{ghost1}, \eq{ghost0} as they are. 
A bit of algebra shows that the amplitude is modified as
\bea
A = {\G(\tilde{s})\G(t)\over \G(1+\tilde{s}+t)} (\tilde{s} a_1 - t a_2),
\label{2pampl}
\eea
where $\tilde{s} = 2 G^{\m\n} (p_\parallel)_\m (p_\parallel)_\n$, 
while the definition of $t$ is the same as the one in Section 3.1.
The polarization dependent part $a_1$ and $a_2$ are 
computed to be
\be
\begin{array}{rcl}
a_1 &=& \Tr(\e_1 D_+) p_1\e_2p_1 - p_1\e_2 D_+ \e_1 p_2 
- p_1 \e_2 \e_1^T D_- p_1 - p_1 \e_2^T \e_1 D_+ p_1 
+ \{ 1 \leftrightarrow 2 \}\cr
&&-p_1\e_2\e_1^T p_2 - p_1 \e_2^T \e_1 p_2 
 - \tilde{s} \Tr(\e_1\e_2^T) ~, 
\cr
a_2 &=& \Tr(\e_1D_+)(p_1\e_2D_+p_2 + p_2D_+\e_2 p_1 
  + p_2 D_+ \e_2 D_+ p_2) \cr
&& +p_1D_+\e_1 D_+ \e_2 D_+ p_2 + \{ 1 \leftrightarrow 2 \} \cr
&& + p_1D_+\e_1\e_2^TD_+p_2 + p_1D_+\e_1^T\e_2D_+p_2 
+ \tilde{s}\Tr(\e_1 D_+ \e_2 D_+) -\tilde{s}\Tr(\e_1\e_2^T) \cr
&& -(\tilde{s}+t) \Tr(\e_1D_+)\Tr(\e_2D_+) ~. 
\end{array}
\label{polar2}
\ee
To keep the notations simple, we introduced 
\be
D_{\pm}^{\m\n} \equiv
D^{\m\n} + 2\Delta^{\m\n} \pm 2\Theta^{\m\n} ~, 
\ee
and
\begin{equation}
\Delta^{\mu \nu} = G^{\mu \nu }-\eta^{\mu \nu} ~.
\end{equation}
We note that $\tilde{s}$ is the $s$-channel momentum transfer 
computed with respect to the open string metric $G^{\mu \nu}$.  
When we push one of the bulk closed string vertex toward the open
string boundary, we approach the $s$-channel factorization
limit where we expect to observe intermediate
open string states.  As was explained in Ref.~\cite{witten},
these open string intermediate states feel the open string
metric $G^{\mu \nu}$ instead of $\eta^{\mu \nu}$, providing
an explanation for the new definition of $\tilde{s}$.  
In the large $B$ limit, therefore, the kinetic
energy along the directions parallel to the two-cycle
along which $B$ is turned on becomes 
negligible, being suppressed by $1/(1+B^2 )$ factor (see also
\cite{bigasuss}).

\subsection{Massless $t$-channel poles}

The modifications due to the non-vanishing $B$ field
in Eqs.~(\ref{polar2}) when compared to Eqs.~(\ref{polar1})
become more transparent when we consider the leading
$t$-channel poles.  Essentially, the behavior of the
string amplitudes around massless $t$-channel poles
contains informations on the long range background
fields.  When expanded around massless $t$-channel 
poles, the scattering amplitude (\ref{2pampl}) 
reduces to
\be
A \sim {1\over t} a_1 + \CO(1) \equiv 
{1\over t} \bA + \CO(1) ~.
\label{answer1}
\ee
The bulk massless string states are gravitons,
$B$ fields and the dilaton.  Therefore, for the two-point
scatterings, there are six possible combinations
of external closed string states.  Among these,
we write down below five possible combinations
by explicitly plugging in the polarization states
into Eqs.~(\ref{polar2}):
\be
\begin{array}{rcl}
\bA(B,\phi) &=& -8p_1 \Theta\e_1 p_2  ~,
\cr
\bA(B,h) &=& 2\{ \Tr(\e_1\Theta)p_2\e_1p_2 
-2p_2\e_1\Theta\e_2p_1 + 2p_2\e_1\e_2\Theta p_2 -2p_1\e_2\e_1\Theta p_1 \} , 
\cr 
\bA(\phi,\phi) &=& 8\tilde{s}, 
\cr  
\bA(\phi, h) &=& 2(p-3)p_1\e_2 p_1 + 2\Tr(\Delta)p_1\e_2p_1, \cr
\bA(B,B) &=& 
-2 p_1\e_2(D+2\Delta)\e_1p_2 + 2p_1\e_2\e_1(D+2\Delta)p_1 \cr
&&+ 2p_2\e_1\e_2(D+2\Delta)p_2 +2p_1\e_2\e_1p_2 
 - \tilde{s}\Tr(\e_1\e_2) ~. 
\end{array}
\label{answer}
\ee
The arguments of $\bA(x,y)$ denote the two external states
$x$ and $y$.  The notable amplitudes are $\bA(B,\phi)$
and $\bA(B,h)$, which vanish when the 
background $B$ field is set to zero \cite{klebanov}. 
>From the supergravity side, massless $t$-pole string 
amplitudes should
be recovered by considering the three-point interactions
through which the long range background field affects
the external states.  An inspection of the low energy supergravity 
action shows that
the possible three-point interactions involving $B$ fields 
are of the type $B$-$B$-graviton or $B$-$B$-dilaton.  Therefore,
the non-vanishing amplitudes $\bA(B,\phi)$ and $\bA(B,h)$
imply that there exists non-trivial long range background
NS-NS two-form gauge field.  When a constant $B$
field is turned on on the brane world-volume, one might
try to gauge it away to zero.  However, for the directions
parallel to the D$p$-branes, we can not gauge it away, since
it simultaneously involves the transformation of the
world-volume $U(1)$ gauge field.  The seemingly trivial 
constant $B$ on the world-volume induces the
non-trivial long range background $B$ fields.


\section{Comparison with Supergravity}

In this section, we do the tree-level supergravity 
calculation of the two-point scatterings in the 
background geometries proposed in Refs.~\cite{aki,maldaruss}.
The main result is that the leading $t$-pole string amplitudes 
computed in Section 3 are identical to the long-range
supergravity tree amplitudes.  
  
\subsection{Supergravity background with or without $B$ field}

The NS-NS sector of the low energy effective action for Type 
II strings in ten dimensions reads, in Einstein frame,  
\be
\label{leea}
S = {1\over 2\kappa^2} \int d^{10}x \sqrt{-g} 
\left\{ R - \half (\del\phi)^2 - {1\over 12} e^{-\phi} H^2 \right\}.
\ee
The solutions representing a stack of parallel $N$ D$p$-branes 
are well-known and take the simplest form in the string frame: 
\bea
ds^2 &=& H^{-1/2} (-dt^2 + \cdots + dx_p^2) 
+ H^{1/2} (dx_{p+1}^2 + \cdots + dx_9^2) ~, \\
e^\phi &=& H^{(3-p)/4} ~, \  \
H \equiv 1 +F \equiv 1 + (R_p/r)^{7-p} ~.
\eea
Recall that the Einstein metric and the string metric is related by
$ds_E^2 = e^{-\phi/2} ds^2$.  Recently, the authors of
Refs.~\cite{aki, maldaruss} showed how to incorporate
the effect of the $B$ field background: 
\bea
ds^2 &=& 
H^{-1/2} \left\{-dt^2 + \cdots + dx_{p-2k}^2 
+ \sum_{i=1}^k N_i (dy_{2i-1}^2 + dy_{2i}^2) \right\}
+ H^{1/2} dx_\perp^2 ~, \label{1sugrasol} \\
B_i &=& H^{-1} N_i \tan\th_i ~, \\
e^\phi &=& H^{(3-p)/4} \prod_{i=1}^k N_i^{1/2} ~ , \  \
N_i^{-1} \equiv \cos^2\th_i + H^{-1} \sin^2\th_i ~.
\label{sugrasol}
\eea
Their derivation is based on the chain of T-duality
arguments suggested by \cite{doughull}.  Actually,
the solutions of \cite{aki} are related to the 
solutions of \cite{maldaruss} by a $B$-dependent
rescaling of the $y$-coordinates.  In our string
calculations in Section 2, we chose $g_{\mu \nu}
= \eta_{\mu \nu}$ and $\tan \theta_i = B_i$. 
We note that the solutions (\ref{sugrasol}) are written
in such a coordinate system that $e^{\phi} \rightarrow 1$,
$ds^2 \rightarrow \eta_{\mu \nu} dx^{\mu} dx^{\nu}$
and $B_i \rightarrow \tan \theta_i$ near the asymptotic
spatial infinity.  Consistent with Section 2, there
exist non-vanishing R-R background fields for 
the D$p$-branes (proportional to
$\cos \theta_i$) and for the D$(p-2)$-branes 
(proportional to $\sin \theta_i$) \cite{maldaruss},
which we do not consider in this paper\footnote{To
check the consistency of the background R-R gauge
fields, we need to consider the two-point scatterings 
between NS-NS fields and R-R fields.}.

\subsection{Tree-level supergravity scatterings}

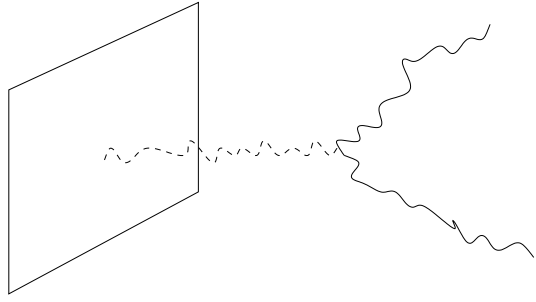
\fig{200pt}{sugra}{Typical supergravity $t$-channel scatterings.
An incoming particle is scattered by the background fields to
an outgoing particle}

We perform the analysis for the long-distance tree-level
supergravity scatterings.  Adding the background
noncommutative D-branes to our problem is tantamount
to adding a source term of the type \cite{klebanov}
\be
S_{source} = \int d^{10}x \sqrt{-g} 
\{S^{\m\n}_h h_{\m\n} + S_{\phi}\phi + S^{\m\n}_B B_{\m\n} \} ~,
\label{source}
\ee
where $h$, $\phi$ and $B$ represent the fluctuations
of graviton (around the flat background), dilaton and
the NS-NS two-form gauge field, respectively.  For source-probe
type scatterings shown in Fig.~2, there are six possible
external state combinations involving massless
NS-NS particles.  One can read off the 
three-point interaction vertices by expanding the 
low energy effective action \eq{leea} in flat spacetime background, and
we find that there are four types of such interactions:
\be
\begin{array}{rcl}
4 V(\phi,\phi,h) &=& 2 p_1\e_3 p_2 - p_1p_2 \Tr(\e_3) ~, \cr
4 V(B,B,\phi) &=&  2p_1\e_2\e_1p_2 - p_1p_2 \Tr(\e_1\e_2) ~, \cr
4 V(B,B, h) &=&  
-p_1\e_3p_2 \Tr(\e_1\e_2) - 2 p_1p_2 \Tr(\e_1\e_3\e_2) \cr
&&+2 p_1\e_3\e_2\e_1 p_2 + 2 p_2 \e_3 \e_1 \e_2 p_1 +2 p_2\e_1\e_3\e_2p_1 \cr
&&+\Tr(\e_3)\left[-p_1\e_2\e_1p_2 
 + \half p_1p_2 \Tr(\e_1\e_2) \right] ~,
\end{array}
\label{vertex}
\ee
and $V(h,h,h)$.  For the scatterings 
shown in Fig.~2, the leading non-trivial
($r$-dependent) background fields are nothing but
\[
  {\rm background~field~of~order~}\frac{1}{r^{7-p}}
  = {\rm source} \times {\rm propagator} ~.
\]
Therefore, we can replace the $t$-channel exchanged particle
part of the diagram in Fig.~2 with its classical background
field.  By expanding the supergravity solutions of
Refs.~\cite{aki} and \cite{maldaruss}, i.e., 
Eqs.~(\ref{1sugrasol})-(\ref{sugrasol}), the 
leading order background fields in Einstein
frame are computed to be:
\be
\begin{array}{rcl}
ds_E^2 &=& ds_{\mbox{flat}}^2 \{1 +{1\over 8}F\Tr(\Delta)\} 
- F \Delta^{\m\n}dx^\m dx^\n + \CO(F^2),  \cr
B_i &=& \tan\th_i + F \Theta_i + \CO(F^2), \cr
\phi &=& -{1\over 4}F(p-3 + \Tr(\Delta)) + \CO(F^2) ~,
\end{array}
\label{backgr}
\ee
where $\Theta_i$ and $\Delta$ are defined in Sec.~2 and
Sec.~3.2, respectively.  We compute $(B , \phi)$, 
 $(B , h)$,  $(\phi , \phi)$,  $(\phi , h)$ and
 $(B , B)$ scatterings using the vertex (\ref{vertex})
and the background fields (\ref{backgr})\footnote{Though
we have not computed the graviton-graviton scatterings,
it is known from the analysis of \cite{klebanov} that
$(h, \phi )$ and $(B,B)$ scatterings are enough to
uniquely determine the graviton and dilaton
source terms in (\ref{source}).}. 
For each scattering, the relevant interaction vertices are
$V(B,B, \phi)$, $V(B,B,h)$, $V(\phi , \phi , h)$,
$ V(\phi , \phi , h)$, $V(B,B, \phi ) + V(B,B, h)$,
respectively.  The final results are identical to
(\ref{answer1}) and
(\ref{answer}).  In this fashion, our string theory
calculations justify the supergravity solutions
of Refs.~\cite{aki,maldaruss}.
   

\section{Absorption and emission by noncommutative D-branes}

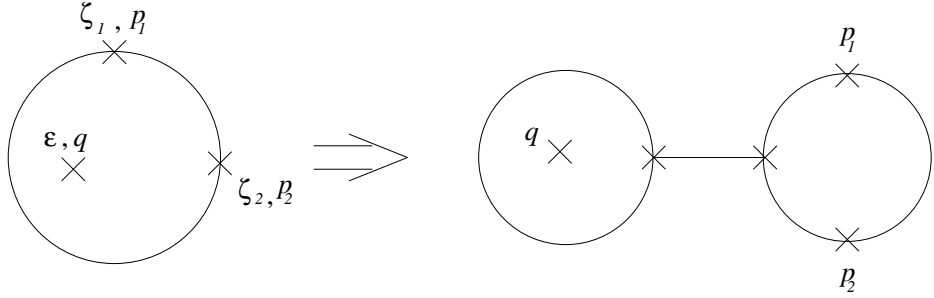
\fig{350pt}{3pt}{Three-point closed string absorption/emission
diagram.  On the right hand side, we show a factorization
limit}

While the two-point scatterings computed in Section 3
can be used to justify the background geometries, the
two-point scatterings do not exhibit the exponential
phase factor used to define the $*$-product \cite{witten,connes}. 
The simplest non-trivial
example showing such phase factor is the string 
emission/absorption diagram from/to noncommutative
D$p$-branes as shown in Fig.~3.  In itself, this amplitude is 
important, for it shows nontrivial modifications of the 
Hawking radiation spectrum from the near-extremal
noncommutative D$p$-brane black holes, when compared
to emissions from commutative D$p$-brane black holes.
Most notably, we find that the pole structure of the 
amplitude changes as we turn on the $B$ 
field, and the exponential phase factors show up. 

\subsection{Review: Absorption/emissions from commutative
D$p$-branes}
 
The amplitude is given by
\be
A = \int_\Sigma  d^2 z \int_{\partial \Sigma} 
dw_1 \int_{\partial \Sigma} dw_2 
\langle V_c(z,\bz;q,\e) V_o(w_1;p_1,\c_1) V_o(w_2;p_2,\c_2)\rangle,
\ee
where the closed and open string vertex operators are
\bea
V_c(z,\bz;q,\e) &=& (\e D)_{\m\n} 
 V_{-1}^\m(q,z) V_{-1}^\n(Dq, \bz) ~,
\\
V_o(w;p,\c) &=& \c_\m 
\{ \partial X^\m(w) + 2ip\cdot\psi(w)\psi^\m(w) \} e^{2ip X(w)},
\eea
respectively.   The kinematics of this scattering
allows only one kinematic invariant $t$ defined as
$t = -2 p_1 \cdot p_2$.  The open string momenta
$p_1$ and $p_2$  are restricted to lie along the
D-brane world-volume, and the momentum is conserved
along the directions parallel to the branes
\begin{equation}
 p_1 + p_2 + q_{\parallel} = 0 ~.
\label{conserv}
\end{equation}
Again, fixing the $SL(2,\IR)$ invariance and performing the 
integral, one finds
\bea
A = {\G(1-2t)\over \G(1-t)^2} K,
\label{comab}
\eea
where the kinematic factor is given by
\be
\begin{array}{rcl}
K &=&
\Tr(\e D)(\c_1 q)(\c_2 D q) + 4(p_1\e p_2)(\c_1\c_2)
+2(\c_2 p_1) \{ \c_1\e D q + q D \e D \c_1 \} \cr
&&-4 (\c_2 q) (\c_1 \e p_1) -4 (\c_2 D q)(p_1\e D \c_2) 
+ \{ 1 \leftrightarrow 2 \} \cr
&& +t \{\Tr(\e D) \c_1\c_2 -2\c_1\e D\c_2 - 2 \c_2\e D\c_1\}.
\end{array}
\ee
It was noted in Ref. \cite{klebanov} that this amplitude has poles 
for half integer values of $t$, but has {\em zeros} for 
integer values of $t$, summarized by  
a sort of ``$\IZ_2$ selection rule''.
In particular, there is no massless pole. 
As we will see shortly, the situation 
drastically changes as we turn on the $B$ field.

\subsection{Turning on the $B$ field}

We can include the effect of the background $B$ field by 
placing the matrix $R$ 
at appropriate places, as was done in Section 2.  The 
closed string vertex operators change in the same way as
in Eqs.~(\ref{svert1}) and (\ref{svert2}) 
and the open string vertex operators become
\be
V_o(w; Mp, M\c),\  \ M \equiv \half(R+R^T) ~.
\ee
Under these modifications, again paying attention to
the fixing of the $SL(2,\IR)$ invariance, we obtain the
following amplitude:
\be
\label{3ptamp}
A = {\G(1-2t) \over \G(1-t-\d) \G(1-t +\d)}
\left( \tilde{K} - {\d^2\over t} a \right) ~,
\ee
where we introduce
\be 
a = \Tr(\e D_+) \c_1  \c_2 ~ ,\  \ 
 \d = 2p_1\Theta p_2 ~ .
\ee
The polarization dependent quantity $\tilde{K}$ 
is obtained from $K$ by the following rules: 
First, $\Tr(\e D)$ is replaced by $\Tr(\e D_+)$. 
Second, contractions of any two of the open string quantities 
($p_1, p_2, \c_1, \c_2$) are made with respect to the open string metric 
$G^{\m\n}$.  Accordingly, for example, $t$ is now defined
as $t = -2 p_{1 \mu} G^{\mu \nu} p_{2 \nu}$.
Third, contractions of two closed string quantities $(\e, q)$
do not change. Finally, for contractions of an open string quantity and 
a closed string quantity, we insert $(G+\Theta)$. For example,
\be
\c q \rightarrow \c_\m (G+\Theta)^{\m\n} q_\n.
\ee
The second and third rules are natural, since $G^{\mu \nu}$
is felt by open strings and the $\eta^{\mu \nu}$ is felt
by closed strings.    

An important feature of the amplitude \eq{3ptamp} is that its pole
structures are distinctively different from the commutative
case, Eq.~(\ref{comab}).  We first consider a non-zero value of
$t$.  When $\delta$ is neither integral
nor half-integral, the amplitude Eq.~(\ref{3ptamp}) has
poles for both integral and half-integral values of
$t$.  When $t = m \pm \delta$ where $m= 1,2, \cdots$,
it has zeros.  When $\delta$ is integral, the usual
commutative $\IZ_2$ selection
rule applies: the amplitude has zeros for integral
values of $t$ and poles for half-integral values of $t$.
When $\delta$ is half-integral, the situation reverses
itself: the amplitude has zeros for half-integral values
of $t$ and poles for integral values of $t$.  
Near $t=0$, we can use the 
formula
\be
\G(1+x) \G(1-x) = {\pi x \over \sin\pi x}
\label{t1}
\ee
to find
\be
A \sim {1\over t} \bA,\  \  
\bA \sim \d \sin( \pi \d) \sim \d (e^{i\pi\d} - e^{-i\pi\d}). 
\label{t2}
\ee

Eqs.~(\ref{t1}) and (\ref{t2}) have further implication.
By repeated use of $\Gamma (1 + x) = x \Gamma (x)$ and
Eq.~(\ref{t1}), the three-point amplitude Eq.~(\ref{3ptamp})
can be written as
\begin{equation}
A = A_0 \exp ( i \pi \delta ) - A_0 \exp ( -i \pi \delta ) ~,
\label{t3}
\end{equation}
where $A_0$ is an odd function under $\delta \rightarrow
- \delta$, for integral values of $t$ including $t = 0$.  
We notice that the exponential factors are exactly
what one expects from the noncommutative Yang-Mills
$*$-product.  In Fig.~3, there are two possible 
factorization channels, one with $p_1$ on top (the shown
figure) and another with $p_2$ on top (the flipped
figure which is not shown).  The amplitude (\ref{t3}),
which is an even function under  $\delta \rightarrow
- \delta$ can 
be thought of as the sum of contributions from these 
two diagrams.
The exponential factors, following \cite{witten} for the
three open string insertions,
will be 
\begin{equation}
 p_1 \Theta p_2 + p_1 \Theta q_{\parallel} 
 + p_2 \Theta q_{\parallel} 
 = p_1 \Theta p_2 - q_{\parallel} \Theta 
  q_{\parallel}  = p_1 \Theta p_2  ~,
\end{equation}
where we used the momentum conservation 
(\ref{conserv}) along the
parallel directions to the branes.  Again using the
same momentum conservation, we observe that $\delta$
can be considered as being linearly proportional to
the external momentum $q_{\parallel}$ and does not
vanish as long as $q_{\parallel}$ is not zero.

The appearance of the $*$-product phase factor for
the external closed strings might seem peculiar: for 
the usual non-commutative Yang-Mills theory (i.e.,
essentially the pure open strings), such
phase factors come from
the planar diagrams where the external legs are
attatched to the fundamental particles.  Meanwhile,
closed strings are described by composite 
operators\footnote{We thank the referee for pointing
out this issue.}.  
One way to understand
this peculiarity is to note that the phase factors
in Eq.~(\ref{t3}) show up in the factorization 
limit of the type shown in Fig.~3.  The factorization
limit of Fig.~3 is similar to the $s$-channel 
factorization limit of two-point scatterings shown
in Fig.~1, where the closed string vertex operator
approaches the open string boundaries.  In this
limit, the string absorption diagram contains a factor 
that corresponds to the three-point (open string) 
insertions along the
open string boundaries, as shown in Fig.~3.  This
factor is responsible for the $*$-product phase
factor.  A related issue is to compute the string
absorption amplitudes via classical supergravity
analysis along the line of, for example, 
\cite{klebanov1}.  In the commutative 
case \cite{klebanov1}, it is
known that string calculations are reproduced by
the classical supergravity analysis.  It is, however,
unclear whether such classical supergravity analysis
can reproduce the $*$-product phase factors shown in
Eq.~(\ref{t3}).  Just like the $s$-channel limit
of two-point scatterings, the $*$-product phase
factor gets produced when the closed string vertex
operator gets pushed very close to the open string
boundaries.  Unlike the $t$-channel limit of two-point
scatterings, this is the limit where one has reasonable
doubts about the validity of the perturbative
classical supergravity.  Even if it is a very interesting 
issue to see whether we can capture the $*$-product 
structure from the supergravity analysis, it is thus
beyond the scope of this paper.  We note that the
exponential phase $\delta$ of Eq.~(\ref{t3}) is taken 
to be vanishingly small in the supergravity 
limit \cite{klebanov1}.

\section{Discussions}

It is remarkable that the simple imposition of
the boundary conditions as in Section 2 allows
us to study rather intricate D$p$-D$(p-2)$ bound
states.  Furthermore, the modifications that
occur due to the constant $B_{\mu \nu}$ in the
string calculations are straightforwardly
accomplished by placing the matrix $R$ (or $R^T$)
of Section 2 at appropriate places of computations.  
There is an immediate technical 
generalization to the
analysis presented in Section 2: one would like 
to extend the disk
diagram analysis to the annulus diagrams,
incorporating the open string loop effects.

The closed string absorption/emission amplitudes
computed in Section 5 show some novel features,
such as the occurrence of the exponential phase
factors depending linearly on the external closed
string momentum.  Their pole and zero structures
are also drastically modified as we turn on the
$B$ field.  An interesting issue is
to capture some of these features through the
(possibly non-perturbative)
supergravity absorption calculations along the
line of, for example, Ref.~\cite{klebanov1}.
This kind of comparison in fact gave some of 
motivations for the AdS/CFT correspondence 
conjecture for commutative D-branes \cite{adscft}.  
In the context of noncommutative D-branes, the 
situation is much more subtle,
for example, because the D$p$-branes seemingly
turn into D$(p-2)$-branes in the large $B$
limit.  

It is amusing to compare our calculation of the
correlation functions to the gravitational back-reactions
considered in Ref.~\cite{kvv}.  The three-point amplitudes
computed in Section 5 represent the absorption of
NS-NS matter fields into noncommutative 
D$p$-branes.  As analyzed in \cite{kvv}, when
almost light-like matter falls into a black hole,
due to the energy conservation, the horizon radius
gets increased; this effect translates to the shift
of the horizon along the incoming null-direction
and can thus be represented by the exponential
phase proportional to the incoming momentum, or
the shift operator.  The three-point amplitudes show
the similar exponential phase factor linearly
proportional to the external momentum of the
absorbed matter.  The formal similarity might go
further; the four-point amplitudes of type where
there are two closed string vertex insertions in
the bulk and two boundary open string vertex insertions
has a factorization channel where two closed
string vertices approach the open string boundary.
In this limit, at least, it apparently
appears possible to expect
an exponential phase factor that depends
quadratically on the external momenta, which,
if exists, would
correspond to the space-time noncommutativity
induced by the world-volume noncommutativity. 
This type of four-point amplitudes represents the
interaction between the emitted Hawking radiation
and the absorbed matters.  In Ref.~\cite{kvv},
there are similar interaction effects summarized by
the following exchange algebra, which 
corresponds to a version of space-time 
noncommutativity:
\begin{equation}
\phi_{in} (p_{in}  ) \phi_{out} ( p_{out} )  
  = e^{ i  \kappa p_{in} \cdot p_{out}} 
    \phi_{out} (p_{out} ) \phi_{in} (p_{in} ) ,
\label{exalg1}
\end{equation}   
where $\phi_{in}$ represents the absorbed matter
and $\phi_{out}$ represents the emitted Hawking
radiation.  In Eq.~(\ref{exalg1}), we notice
an exponential phase factor that depends 
quadratically on the external momenta.  It remains to be 
seen whether there is any reason why the space-time 
noncommutativity in Eq.~(\ref{exalg1}) 
(due to the graviton exchanges near the black
hole horizon) is formally 
similar to the possible space-time noncommutativity
induced by the background $B$ field.

\section*{Acknowledgements}

We would like to thank Hyeonjoon Shin for useful discussions.
C.-Y. L. was supported in part by Korea Research Foundation, 
Interdisciplinary Research
Project No. D00001 and BSRI Program 1998-015-D00073.

\end{document}